\documentclass[aps]{revtex4}

\usepackage{graphics,epsfig}
\usepackage{amsmath}
\usepackage{amssymb}
\usepackage{amsfonts}
\usepackage{rotate}
\usepackage{epsfig}

\begin{document}

\title{A network-based prediction of retail stores commercial 
categories and optimal locations}

\author{Pablo Jensen} 

\email{pablo.jensen@ens-lyon.fr}
\affiliation{Laboratoire de Physique, CNRS UMR 5672, Ecole Normale
  Sup\'erieure de Lyon, 46 All\'ee d'Italie, 69364 Lyon Cedex 07,
  France \\ Laboratoire d'Economie des Transports, CNRS UMR 5593,
  ISH-Universit\'e Lyon-2, 14, Av. Berthelot, 69007 Lyon, France}
\date{\today}

\begin{abstract}
  I study the spatial organization of retail commercial activities.
  These are organized in a network comprising ``anti-links'', i.e.
  links of negative weight. From pure location data, network analysis
  leads to a community structure that closely follows the commercial
  classification of the US Department of Labor. The interaction
  network allows to build a 'quality' index of optimal location niches
  for stores, which has been empirically tested.
\end{abstract}


\maketitle

\begin{figure}
\centerline{
\epsfxsize=6cm
\epsfbox{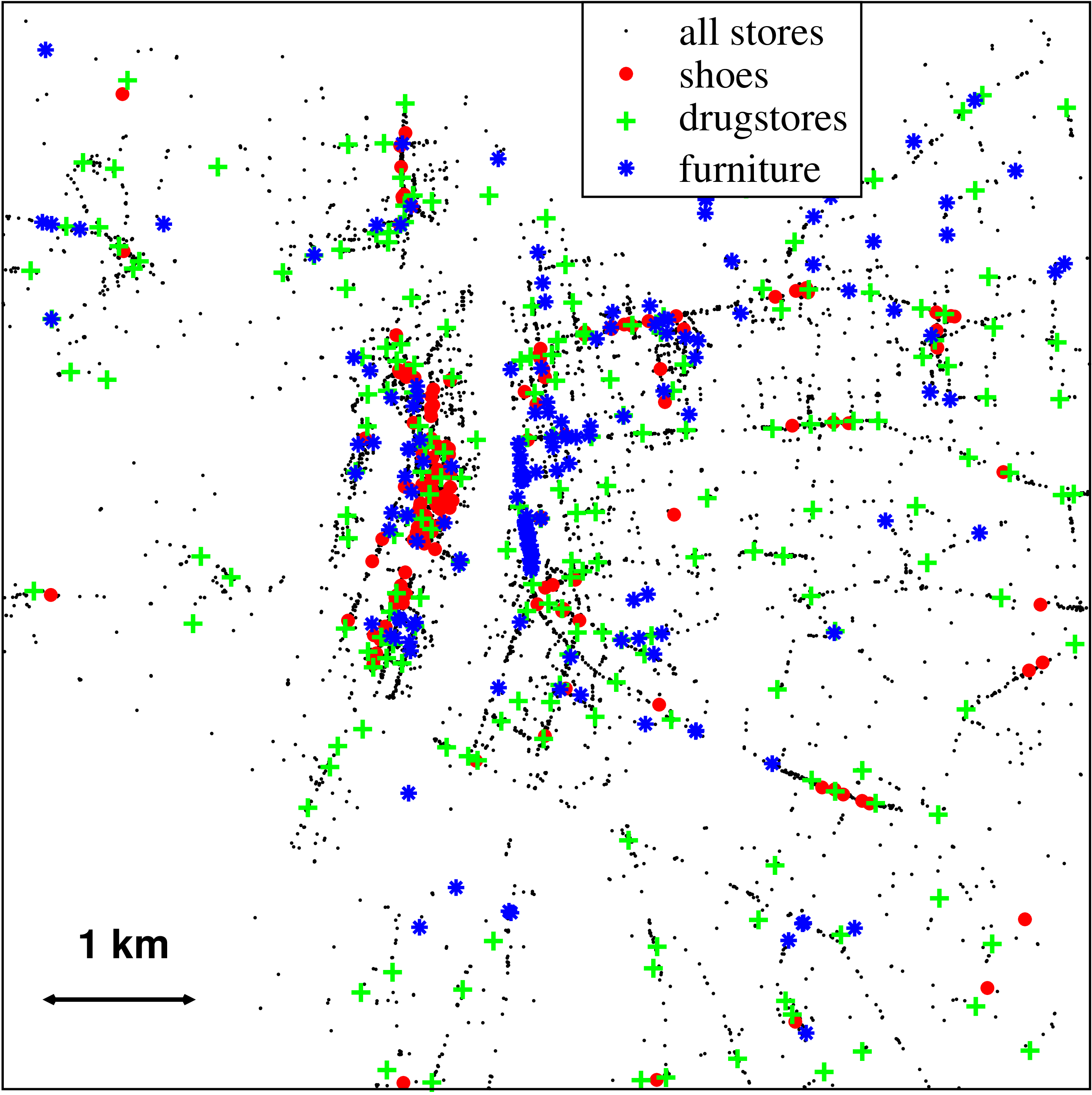}}
\caption{\vspace{.3cm} 
(Color online) Map of Lyon showing the location of all the retail
stores, shoe stores, furniture dealers and drugstores}
\label{carte}
\end{figure}

Walking in any big city reveals the extreme diversity of retail store
location patterns. Fig. \ \ref{carte} shows a map of the city of Lyon
(France) including all the drugstores, shoes stores and furniture
stores. A qualitative commercial organisation is visible in this map :
shoe stores aggregate at the town shopping center, while furniture
stores are partially dispersed on secondary poles and drugstores are
strongly dispersed across the whole town.  Understanding this kind of
features and, more generally, the commercial logics of the spatial
distribution of retail stores, seems a complex task. Many factors
could play important roles, arising from the distincts characteristics
of the stores or the location sites.  Stores differ by product sold,
surface, number of employees, total sales per month or inauguration
date.  Locations differ by price of space, local consumer
characteristics, visibility (corner locations for example) or
accessibility. Only by taking into account most of these complex
features of retail world can we hope to understand the logics of store
commercial strategies, let alone finding potentially interesting
locations for new businesses.

Here I show that location data alone suffices to reveal many important
facts about the commercial organisation of retail trade \cite{data}.
First, I quantify the interactions among activities using network
analysis. I find a few homogeneous commercial categories for the 55
trades in Lyon. These groups closely match the usual commercial
categories : personal services, home furniture, food stores and
apparel stores.  Second, I introduce a quality indicator for the
location of a given activity and empirically test its relevance. I
stress that these results are obtained from a mathematical analysis of
solely {\it location} data.  This supports the importance of business
location for retailers, a point that is intuitively well-known in the
field, and summarized by the retailing ``mantra'' : {\it the three
  points that matter most in a retailer's world are : location,
  location and ...  location}.

\vspace{.5cm}
{\bf Finding meaningful commercial categories}
\vspace{.2cm}

To analyze in detail the interactions of stores of different trades, I
start from the spatial pair correlations. These functions are used to
reveal store-store interactions, as atom-atom interactions are deduced
from atomic distribution functions in materials science \cite{pair}.
Tools from that discipline cannot be used directly, though, because
there is no underlying crystalline substrate to define a reference
distribution.  Neither is a homogeneous space appropriate, since the
density of consumers is not uniform and some town areas cannot host
stores, as is clearly seen in the blank spaces of the map (due to the
presence of rivers, parks, or residential spaces defined by town
regulations).

A clever idea proposed by G. Duranton and H. G. Overman
\cite{duranton} is to take as reference a random distribution of
stores located on the array of {\it all existing} sites (black dots in
Fig. \ \ref{carte}).  This is the best way to take into account
automatically the geographical peculiarities of each town. I then use
the ``M'' index \cite{puech} to quantify the spatial interactions
between categories of stores.  The definition of $M_{AB}$ at a given
distance {\it r} is straightforward : draw a disk of radius {\it r}
around each store of category A, count the total number of stores
($n_{tot}$), the number of B stores ($n_B$) and compare the ratio $n_B
/ n_{tot}$ to the average ratio $N_B / N_{tot}$ where capital N refer
to the total number of stores in town. If this ratio, averaged over
all A stores, is larger than 1, this means that A ``attracts'' B,
otherwise that there is repulsion between these two activities
\cite{note_ave}. To ascertain the statistical significance of the
repulsion or attraction, I have simulated 800 random distributions of
$n_B$ stores on all possible sites, calculating for each distribution
the $n_B / n_{tot}$ ratio around the same A locations.  This gives the
statistical fluctuations and allows to calculate how many times the
random ratio deviates from 1 as much as the real one.  I assume that
if there are less than $3\%$ random runs that deviate more than the
real one, the result is significant ($97\%$ confidence interval). I
have chosen $r=100m$ as this represents a typical distance a customer
accepts to walk to visit different stores \cite{note_dist}

I can now define a network structure of retail stores.  Nodes are
defined as the 55 retail activities (Table I). The weighted
\cite{weight} links are given by $a_{AB} \equiv \log(M_{AB})$, which
reveal the spatial attraction or repulsion between activities A and B
\cite{note_stat}.  This retail network represents the first a social
network with quantified ``anti-links'', i.e.  repulsive links between
nodes \cite{repulsion}.  The anti-links add to the usual (positive)
links and to the absence of any significant link, forming an essential
part of the network. If only positive links are used, the analysis
leads to different results, which are less satisfactory (see below).

To divide the store network into communities, I adapt the ``Potts''
algorithm \cite{potts}. This algorithm identifies the store types as
magnetic spins and groups them in several homogeneous magnetic domains
to minimize the system energy.  Anti-links can then be interpreted as
anti-ferromagnetic interactions between the spins.  Therefore, this
algorithm naturally groups the activities that attract each other, and
places trades that repel into different groups. A natural definition
\cite{potts,radicchi} of the satisfaction ($-1 \leq s_i \leq 1$) of
site $i$ to belong to group $\sigma_i$ is :

\begin{equation}
s_i \equiv {{\sum_{j \neq i} a_{ij} \pi_{\sigma_i \sigma_j}} \over {\sum_{j \neq i} |a_{ij}| }}
\label{s}
\end{equation}

where $\pi_{\sigma_i \sigma_j} \equiv 1$ if $\sigma_i = \sigma_j$ and
$\pi_{\sigma_i \sigma_j} \equiv -1$ if $\sigma_i \not= \sigma_j$.

To obtain the group structure, I run a standard simulated annealing
algorithm \cite{sa} to maximize the overall site satisfaction (without
the normalizing denominator) :

\begin{equation}
K \equiv \sum_{i,j = 1,55; i \neq j} a_{ij} \pi_{\sigma_i \sigma_j} 
\label{K}
\end{equation}

Pott's algorithm divides the retail store network into five
homogeneous groups (Table I, note that the number of groups is not
fixed in advance but a variable of the maximisation). This group
division reaches a global satisfaction of $80 \%$ of the maximum K
value and captures more than $90 \%$ of positive interactions inside
groups.  Except for one category (``Repair of shoes''), our groups are
communities in the strong sense of Ref.  \cite{radicchi}.  This means
that the grouping achieves a positive satisfaction for every element
of the group. This is remarkable since hundreds of ``frustrated''
triplets exist \cite{note_frust}. Taking into account only the
positive links and using the modularity algorithm \cite{modularity}
leads to two large communities, whose commercial interpretation is
less clear.

Two arguments ascertain the commercial relevance of this
classification.  First, the grouping closely follows the usual
categories defined in commercial classifications, as the U.S.
Department of Labor Standard Industrial Classification System
\cite{sic} (see Table I). It is remarkable that, starting exclusively
from location data, one can recover most of such a significant
commercial structure. Such a significant classification has also been
found for Brussels and Marseilles stores (to be presented elsewhere),
suggesting the universality of the classification for European towns.
There are only a few exceptions, mostly non-food proximity stores
which belong to the ``Food store'' group or vice-versa. Second, the
different groups are homogeneous in relation to correlation with
population density. The majority of stores from groups 1 and 2 (18 out
of 26) locate according to population density, while most of the
remaining stores (22 out of 29) ignore this characteristic
\cite{note_corr}. Exceptions can be explained by the small number of
stores or the strong heterogeneities \cite{note_hetero} of those
activities.

\vspace{.5cm}
{\bf From interactions to location niches}
\vspace{.2cm}

Thanks to the quantification of retail store interactions, we can
construct a mathematical index to automatically detect promising
locations for retail stores. The basic idea is that a location that
resembles the average location of the actual bakeries might well be a
good location for a new bakery. To characterize the average
environment of activity $i$, we use the average number of neighbor
stores (inside a circle of radius 100 m) of all the activities $j$,
thus obtaining the list of {\it average} $\overline{nei_{ij}}$.  We
then use the network matrix $a_{ij}$ to quantify deviations from this
average. For example, if an environment lacks a bakery (or other shops
that are usually repelled by bakeries), this should increase the
suitability of that location. We then calculate the quality $Q_i(x,y)$
of an environment around (x,y) for an activity $i$ as :

\begin{equation}
Q_i(x,y) \equiv \sum_{j = 1,55} a_{ij} (nei_{ij}(x,y)-\overline{nei_{ij}})
\label{quality}
\end{equation}

where $nei_{ij}(x,y)$ represents the number of neighbor stores around
x,y. To calculate the location quality for an existing store, one
removes it from town and calculates $Q$ at its location.

\begin{figure}
\centerline{(a) \hspace{.1cm}
\epsfxsize=4cm
\epsfbox{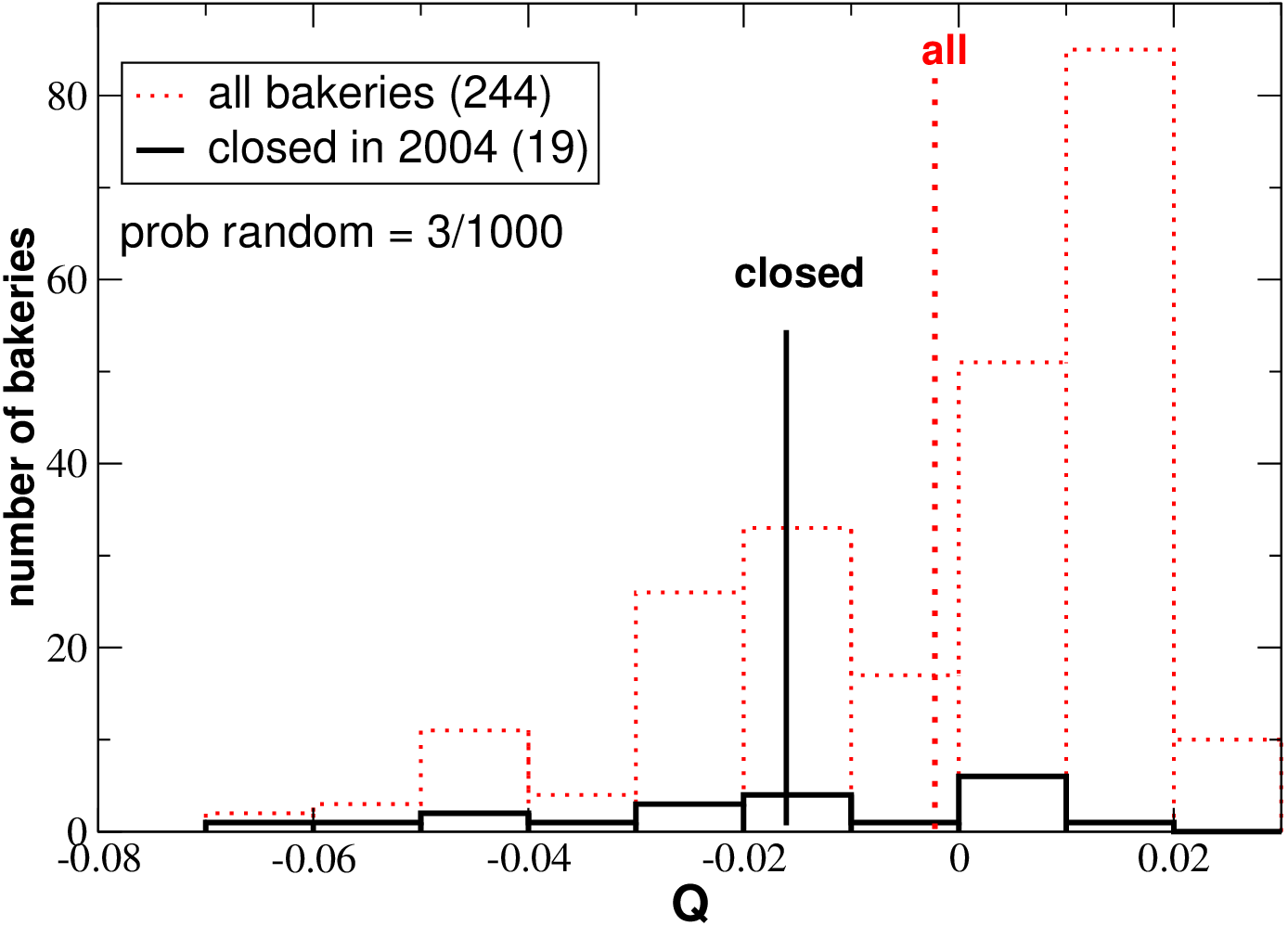}
(b) \hspace{.1cm}
\epsfxsize=4cm
\epsfbox{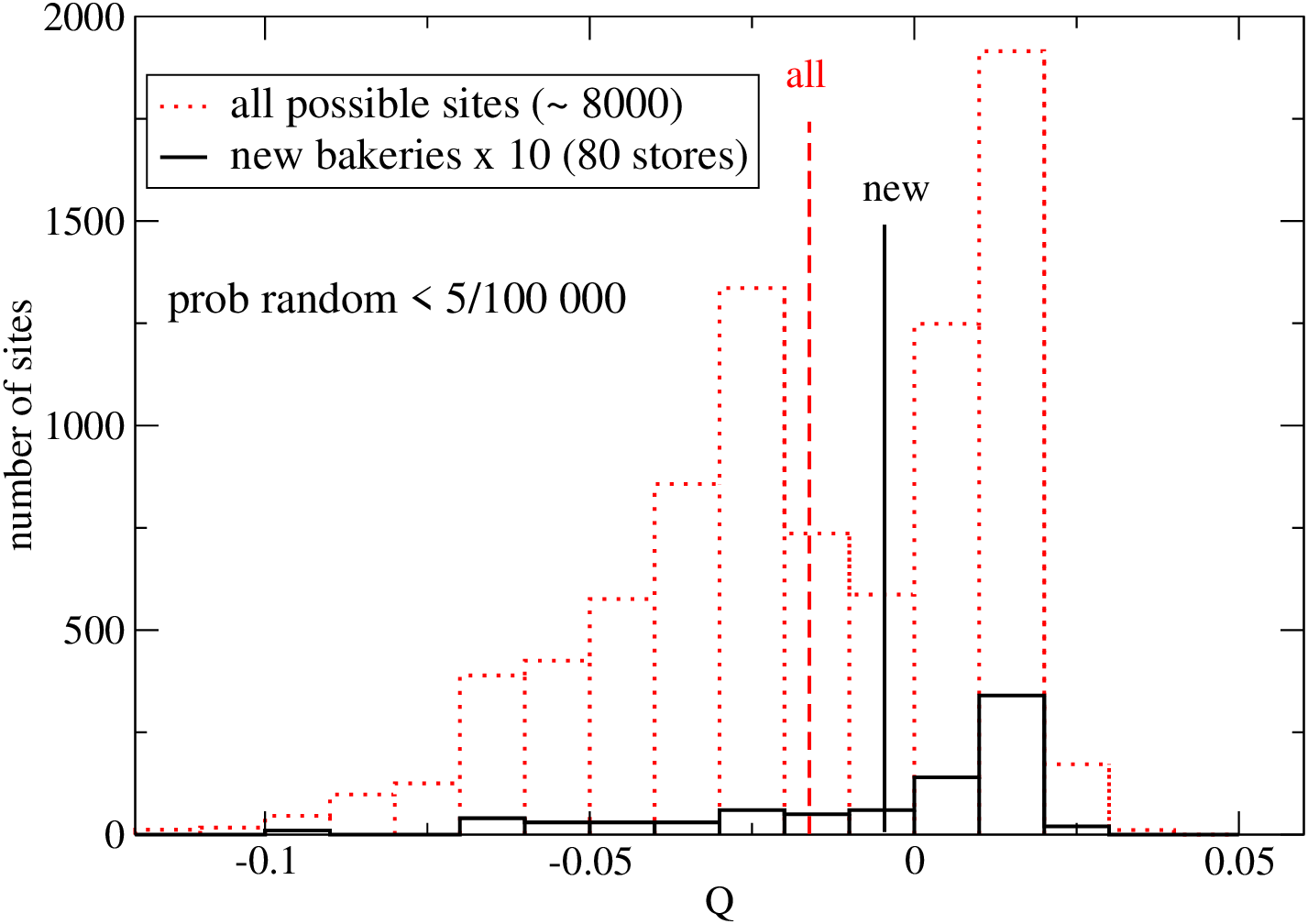}}
\caption{ (Color online) The landscape defined by the quality index is closely
  correlated to the location decisions of bakeries. (a) The 19
  bakeries that closed between 2003 and 2005 had an average quality of
  $-2.2$ x $ 10^{-3}$ to be compared to the average of all bakeries ($4.6$
   x $10^{-3}$), the difference being signifcative with probability
  0.997). Taking into account the small number of closed bakeries and
  the importance of many other factors in the closing decision (family
  problems, bad management...), the sensitivity of the quality index
  is remarkable. (b) Concerning the 80 new bakeries in the 2005
  database (20 truly new, the rest being an improvement of the
  database), their average quality is $-6.8$ x $10^{-4}$, to be compared
  to the average quality of all possible sites in Lyon ($-1.6$ x
  $10^{-2}$), a difference significant with probability higher than
  0.9999).}
\label{0305}
\end{figure}

As often in social contexts, it is difficult to test empirically the
relevance of our quality index. In principle, one should open several
bakeries at different locations and test whether those located at the
``best'' places (as defined by $Q$) are on average more successful.
Since it may be difficult to fund this kind of experiment, I use
location data from two years, 2003 and 2005. It turns out (Fig. \
\ref{0305}) that bakeries closed between these two years are located
on significantly lower quality sites. Inversely, new bakeries (not
present in the 2003 database) do locate preferently on better places
than a random choice would dictate. This stresses the importance of
location for bakeries, and the relevance of the quality here defined
to quantify the interest of each possible site. Possibly, the
correlation would be less satisfactory for retail activities whose
locations are not so critical for commercial success. Practical
applications of $Q$ are under development together with Lyon's Chamber
of Commerce and Industry : advice to newcommers on good locations,
advice to city mayor's on improving commercial opportunities on
specific town sectors.

This study shows that, through locations, the retail world is now
accessible to physicists. This opens many research directions, such as
: are there optimum store distributions, whose overall quality is
higher than the actual one? Can one define store-store interaction
"potentials" by analogy with those used for atomic species? Moreover,
new tools are needed to describe networks containing anti-links,
starting with a basic one : ``how to define a node degree?''.

\vspace{1cm}

\begin{center} {\bf Table I} Retail store groups obtained from Pott's
  algorithm. Our groups closely match the categories of the U.S.
  Department of Labor Standard Industrial Classification (SIC) System
  : group 1 corresponds to Personal Services, 2 to Food stores, 3 to
  Home Furniture, 4 to Apparel and Accessory Stores and 5 to Used
  Merchandise Stores. The columns correspond to : group number,
  activity name, satisfaction, activity concentration (see below),
  median distance travelled by costumers, correlation with population
  density (U stands for uncorrelated, P for Population correlated) and
  finally number of stores of that activity in Lyon. The activity
  concentration $c_{same}$ represents the number of stores located
  nearer than 100 m from another similar store, normalized to the
  number expected from a random distribution. For space reasons, only
  activities with more than 50 stores are shown.
\end{center}

{\it
\begin{tabbing} 
  group\=activity  \hspace{6cm} \= s \hspace{.95cm} \= $c_{same}$
  \hspace{.5cm} \=distance \hspace{.5cm} 
  \= pop corr \hspace{.5cm} \= $N_{stores}$\\ \\
  1\>bookstores and newspapers\>1.00\>1.00\> \>U\>250\\ 
  1\>Repair of electronic household goods\>0.71\>1.00\>1.16\>P\>54\\ 
  1\>make up, beauty treatment\>0.68\>1.00\>1.20\>P\>255\\ 
  1\>hairdressers\>0.67\>0.67\>0.99\>P\>844\\ 
  1\>Power Laundries\>0.66\>1.00\>1.48\>P\>210\\ 
  1\>Drug Stores\>0.55\>0.21\>1.09\>P\>235\\ 
  1\>Bakery (from frozen bread)\>0.54\>0.29\>0.00\>P\>93\\ \\
 
  2\>Other repair of personal goods\>1.00\>1.00\> \>U\>111\\ 
  2\>Photographic Studios\>1.00\>1.00\> \>P\>94\\ 
  2\>delicatessen\>0.91\>1.00\>0.77\>U\>246\\ 
  2\>grocery ( surface $< 120 m^2$)\>0.77\>0.61\>0.00\>P\>294\\ 
  2\>cakes\>0.77\>1.00\>0.35\>P\>99\\ 
  2\>Miscellaneous food stores\>0.75\>2.22\>0.00\>P\>80\\ 
  2\>bread, cakes\>0.70\>1.00\> \>U\>56\\ 
  2\>tobacco products\>0.70\>0.38\> \>P\>162\\ 
  2\>hardware, paints (surface $< 400 m^2$)\>0.69\>1.00\> \>U\>63\\ 
  2\>meat \>0.64\>1.41\>0.86\>P\>244\\ 
  2\>flowers\>0.58\>0.65\>1.52\>P\>200\\ 
  2\>retail bakeries (home made)\>0.47\>0.36\>0.00\>P\>248\\ 
  2\>alcoholic and other beverages\>0.17\>1.00\>0.77\>U\>67\\ \\

  3\>Computer\>1.00\>1.00\>3.07\>P\>251\\ 
  3\>medical and orthopaedic goods\>1.00\>1.00\> \>U\>63\\ 
  3\>Sale and repair of motor vehicles\>1.00\>1.00\>1.68\>P\>285\\ 
  3\>sport, fishing, camping goods\>1.00\>1.00\>2.73\>U\>119\\ 
  3\>Sale of motor vehicle accessories\>0.67\>0.00\>0.00\>U\>54\\ 
  3\>furniture, household articles \>0.62\>3.15\>2.57\>U\>172\\ 
  3\>household appliances\>0.48\>1.00\>3.08\>U\>171\\ 

  4\>cosmetic and toilet articles\>1.00\>2.09\>2.57\>U\>98\\ 
  4\>Jewellery\>1.00\>5.85\>2.77\>U\>230\\ 
  4\>shoes\>1.00\>5.76\>2.43\>U\>178\\ 
  4\>textiles\>1.00\>2.39\>3.87\>U\>103\\ 
  4\>watches, clocks and jewellery\>1.00\>5.02\>2.77\>U\>92\\ 
  4\>clothing\>0.91\>5.10\>3.16\>U\>914\\ 
  4\>tableware\>0.83\>1.96\>2.43\>U\>183\\ 
  4\>opticians\>0.78\>1.98\>1.55\>U\>137\\ 
  4\>Other retail sale in specialized stores\>0.77\>1.51\>2.32\>U\>367\\ 
  4\>Other personal services \>0.41\>1.00\> \>U\>92\\ 
  4\>Repair of boots, shoes \>-0.18\>1.00\> \>U\>77\\ \\
 
  5\>second-hand goods \>0.97\>16.13\>3.52\>U\>410\\ 
  5\>framing, upholstery\>0.81\>1.67\> \>U\>135\\ 

\end{tabbing}
}

\vspace{2cm}

\begin{thebibliography}{100}
\bibitem{data} Christophe Baume and Frederic Miribel (commerce
  chamber, Lyon) have kindly provided extensive location data for 8500
  stores of the city of Lyon.
\bibitem{note_ave} 
One could argue that the average is dominated by the denser regions,
thus eliminating the influence of peripheral areas. This effect
exists, even if it is partially corrected through the ponderation by
the total number of stores. I have tried several other statistical
representation of the relative concentration, such as the mode or the
median, but none performed as well as the average. The median, for
example, fails because most $A$ stores have no $B$ stores around them,
leading to mostly null interaction coefficients.
\bibitem{note_dist} 
Alternatively, one can fully count stores closer than 50 m and linearly
decrease the counting coefficient until 150 m. This leads to similar
results.
\bibitem{note_stat} For a pair interaction to be significant, I demand
  that both $a_{AB}$ and $a_{BA}$ be different from zero, to avoid
  artificial correlations \cite{puech}.  For Lyon's city, I end up
  with 300 significant interactions (roughly $10 \%$ of all possible
  interactions), of which half are repulsive.
\bibitem{note_frust} A frustrated (A,B,C) triplet is one for which A
  attracts B, B attracts C, but A repels C, which is the case for the
  triplet shown in Fig. \ \ref{carte}.
\bibitem{note_hetero} Several retail categories defined by the
  Commerce Chamber are unfortunately heterogeneous : for example,
  ``Meat'' refers to the proximity butcher stores, but also to a big
  commercial pole of casher butchers who attract costumers from far
  away towns. ``Bookstores and newspapers'' refers to big stores
  selling books and CDs as well as to the proximity newspaper
  stand. Instead, bakeries are precisely classified in 4 different
  categories : it is a French commercial structure!
\end{thebibliography}
\end{document}